# Distribution of small dispersive coal dust particles and absorbed radioactive chemical elements in conditions of forced acoustic resonance in iodine air filter at nuclear power plant

Oleg P. Ledenyov and Ivan M. Neklyudov

*Institute of Solid State Physics, Materials Science and Technologies*
*National Scientific Centre Kharkov Institute of Physics and Technology,*
*Academicheskaya 1, Kharkov 61108, Ukraine.*

The physical features of distribution of the small dispersive coal dust particles and the adsorbed radioactive chemical elements and their isotopes in the absorber with the granular filtering medium with the cylindrical coal granules were researched in the case of the intensive air-dust aerosol stream flow through the iodine air filter *(IAF)*. It was shown that, at the certain aerodynamic conditions in the *IAF*, the generation of the acoustic oscillations is possible. It was found that the acoustic oscillations generation results in an appearance of the standing acoustic waves of the air pressure (density) in the *IAF*. In the case of the intensive blow of the air-dust aerosol, it was demonstrated that the standing acoustic waves have some strong influences on both: *1)* the dynamics of small dispersive coal dust particles movement and their accumulation in the *IAF*; *2)* the over-saturation of the cylindrical coal granules by the adsorbed radioactive chemical elements and their isotopes in the regions, where the anti-nodes of the acoustic waves are positioned. Finally, we completed the comparative analysis of the theoretical calculations with the experimental results, obtained for the cases of: *1)* the experimental aerodynamic modeling of physical processes of the absorbed radioactive chemical elements and their isotopes distribution in the *IAF*; and *2)* the gamma-activation spectroscopy analysis of the absorbed radioactive chemical elements and their isotopes distribution in the *IAF*. We made the innovative propositions on the necessary technical modifications with the purpose to improve the *IAF's* technical characteristics and increase it's operational time at the nuclear power plant (*NPP*), going from the completed precise characterization of the *IAF's* parameters at the long term operation.

## Introduction

The iodine air filters *(IAF)* with the granular filtering medium with the cylindrical coal granules are used with the purpose of air filtering from the radioactive chemical elements and their isotopes in the heating ventilation and cooling systems at the nuclear power stations *(NPP)* with various types of nuclear reactors. The *IAF* play an important role by reducing the radioactive chemical elements and isotopes emission in the atmosphere and by protecting the environment from the possible radioactive contamination, allowing the safe operation of the *NPP*. The *IAFs* of the type of *AU-1500* are intended for the long term duration of operation (several years) and continued to be used without the design changes at the *NPPs* in *Ukraine*. At the same time, it was found that there is a considerable number of physical and technical features, which may affect the technical characteristics of the air filtering device. For example, it was found that, at a certain operational stage, the sharp increase of the *IAF's* aerodynamic resistance takes place usually, resulting in the *IAF's* operational failure. In addition, the random unpredictable emissions of the accumulated radioactive chemical elements and their isotopes may occur, having no any concise explanation, based on the *IAF's* design. Initially, it was not possible to foresee all the technical failures during the *IAF's* operation, because some of them are connected with the complex physical processes, which can only be observed in the *IAF* at the application of long term operational stresses. Therefore, the nature of these complex physical processes has to be found and correlated with the observed phenomena.

Let us notice that the research problem on the physical processes in the *IAF*, which operates under the multiple actions by the dust, radiation and aerodynamic loads, influencing the granular filtering medium with the cylindrical coal granules with the porous internal structure, is quite difficult. This research problem is within the scope of the physics of dust mediums. The dust mediums are intensively investigated in the physics of interstellar space [1, 2]; the physics of plasma: researches on the dust crystals, dust ionic-acoustic waves, quantum dust acoustic waves [3 – 8]; and the physics of aerosols [9 – 11] especially. The gases, containing the granular solid state particles, form a special class of the granular gases with their own properties as far as their particles dispersion, energies distribution, temperatures, possibility of nonlinear interaction and structurization is concerned [12 – 14]. At the same time, the properties of filed granular mediums essentially differ from the properties of continuous homogeneous materials in terms of the elastic and deformation properties [15]; the possible types of fluctuations; and the distribution of stress fields at an influence by the external forces stresses [16]. The transposition and structurization of dust masses in such granular filtering mediums are characterized by the possibility of appearance of the small dispersive coal dust particles concentration density maximums in the core of the *IAF*, which were found to exist in the *IAF* models [17, 18]. In the course of research on the *AU-1500 IAF* by the method of gamma-activation analysis, it has been shown [19] that the distributions of a number of the adsorbed radioactive chemical elements and their isotopes are characterized by a few explicit alternate maximums, located in the adsorber core of the *IAF* after its long-term operation. These maximums were positioned along the direction of the air-dust aerosol stream flow in the *IAF*.

In [20], it was shown that spatial arrangements of both the maxima of the small dispersive coal dust particles, which have precipitated from the air - dust aerosol stream, and the maxima of distributed radioactive chemical elements and their isotopes correlate among themselves along the length of adsorber in the *IAF*. The density of accumulation of adsorbed radioactive chemical elements and their isotopes in the dust mass unit exceeds a similar indicator in the granular filtering medium with the cylindrical coal granules in the *IAF*. In the small dispersive coal dust particles masses, which appear at the partial destruction of the coal granules of adsorbent at the aerosol stream flow, the access by the radioactive chemical elements and their isotopes to the nano-dimensional pores as well as the absorption of the radioactive chemical elements and their isotopes in the nano-dimensional pores are significantly increased. It is connected with the fact that the relation between the external surface of a cylindrical coal granule, through which the penetration of the radioactive chemical elements and their isotopes into the cylindrical coal granule with their subsequent capture on the adsorbing internal surface of internal pours takes place, and the volume of a cylindrical coal granule is: $1/d_{GR}$, where $d_{GR}$ is the average size of a cylindrical coal granule. At the same time, in the case of the small dispersive coal dust particles, the value of $1/d_P$ is in $10^2 \div 10^3$ times bigger. The time of diffusional transposition of the radioactive chemical elements and their isotopes from the surface into the depth of a cylindrical coal granule is $\tau \sim d_{GR}^2/<D>$, where $<D>$ is the averaged coefficient of diffusion of radioactive chemical element in the air and in the internal adsorbing surface of a pore in the cylindrical coal granule, because these two physical mechanisms are involved in the considered process. Let us take to the consideration the fact that the quantity of the radioactive chemical elements and their isotopes, penetrating into a small coal dust particle, is proportional to the relative availability of internal adsorbing surface area in a small coal dust particle, as it is considered above, and divided by the time of penetration $J \propto <D>/d^3$ [21]. As the characteristic sizes of both the small dispersive coal dust particles and the cylindrical coal granules differ in $10^2 \div 10^3$ times, hence the adsorption process in the small dispersive coal dust particles is in many times faster, comparing to the cylindrical coal granules. Therefore, the non-uniform distribution of small dispersive coal dust particles results in the non-homogeneous distribution of the accumulated radioactive chemical elements and their isotopes along the length of the *IAF*. This conclusion allows us to interconnect the positions of maxima of distribution of the small dispersive coal dust particles with the positions of maxima of distribution of the adsorbed radioactive chemical elements and their isotopes, however, at the same time, it does not allow us to make the enough substantiated conclusion about the reasons of appearance of such small dispersive coal dust particles maxima in the core of the *IAF*. In [17], it was shown that there may be a possibility of realization of the diffusional mechanism in the processes of the separation and division of the small dispersive coal dust particles by their masses, which is similar to the molecular-level diffusional mechanism in the chromatographic columns. However, this mechanism does not allow us to explain all the physical features of distribution of the small dispersive coal dust masses in the *IAF*; and it describes probably the diffusional dust "tails" in the close proximity to the frontal region of an absorber in the *IAF*. At the same time, the gamma-resonant researches demonstrate that there are the maxima in the distribution of the accumulated radioactive chemical elements and their isotopes in the absorber near to the *IAF's* output, where the small dispersive coal dust particles masses are not present almost, and hence, the absorbed radioactive chemical elements and their isotopes are captured in the cylindrical coal granules mainly [19]. This important research expands the frameworks of our early conducted investigations, and it focuses on the finding of possible influences by the acoustic phenomena on the distribution of the small dispersive coal dust particles masses together with the adsorbed radioactive chemical elements and their isotopes in the granular filtering medium with the cylindrical coal granules at an action by the air – dust aerosol stream in the *IAF*. The forced acoustic oscillations can be excited at the air – dust aerosol stream flow in the non-homogeneous granular filtering medium with the cylindrical coal granules, and they are usually accompanied by the energy dissipation processes. The present research is based on an assumption that the length of the excited standing acoustic waves $\lambda$ is significantly bigger than the characteristic sizes of the cylindrical coal granules, $\lambda \gg d_{GR}$, but it is comparable with the *IAF's* length, $\lambda \sim L$.

# Discussion on nature of acoustic phenomena in iodine air filter

Let's consider the nature of acoustic phenomena in the granular filtering medium of an absorber in the *IAF* in the case, when there is an air pressures difference at the input and output ports, $\delta P$, and hence, the air-dust aerosol stream, *J*, transports the small dispersive coal dust particles in the *IAF*. Let us believe that the *C* is the small dispersive coal dust particles load, which is equal to the relative mass concentration of the small dispersive coal dust particles in the unit of air volume, $C = m_p/m_v$, and the value of *C* is small, $C \ll 1$. As it was shown in [17, 18, 20], the air propagation through the granular filtering medium with the cylindrical coal granules in the *IAF* is accompanied by the process of structurization of the small dispersive coal dust particles masses; and, considering the *IAF's* long time operation cycle, the relative quantity of the accumulated small dispersive coal dust particles can reach up to almost *100 %* of the free volume, existing in the granular filtering medium between the cylindrical coal granules of an absorber in the *IAF*. For example, at the big enough integral small dispersive coal dust masses loads, there is an increased density of the small dispersive coal dust particles masses concentration at the sub-surface layer of an absorber's input in the *IAF*. The small dispersive coal dust particles masses concentration has a distribution dependence with a few maxima of density along the length of an absorber in the *IAF*. The possible theoretical mechanisms of such physical behaviour by the small dispersive coal dust particles masses is still unknown, hence, in the present research project, we would like to investigate the influences by the forced acoustic oscillations, appearing in the granular filtering medium at the air stream flow, on the distribution of both the small dispersive coal dust particles masses and the absorbed chemical elements and their isotopes in the core of an absorber in the *IAF*.

As we reported earlier [17], the process of transposition of the small dispersive coal dust particles by the air stream depends on the several physical forces such as: the force of viscous capture of the small dispersive coal dust particles by the air; the force of inertia in the regions, where the air – dust aerosol stream, flowing around the granules, sharply changes its direction; and the gravitational force; which act on the particles.

In the present research article, we conduct the research on the possible influence by the acoustic oscillations on the distribution of the small dispersive coal dust particles and the adsorbed radioactive chemical elements and their isotopes in the *IAF*. The acoustic oscillations can be generated by the air stream, which flows between the cylindrical coal granules in view of the existing air pressures difference in the *IAF*, which is created by the external air pump.

Let's consider the physical principles of the acoustic oscillations generation during the air stream flow in the *IAF*. In the case of the continuous air stream flow, there is a sharp change of the air stream velocity *V* at the *IAF's* boundaries, and the internal volume can be considered as a tube with the potentials gap, where the magnitude of the internal air stream velocity *V* is much bigger, comparing to the magnitude of the external air stream velocity as shown in Fig. 1.

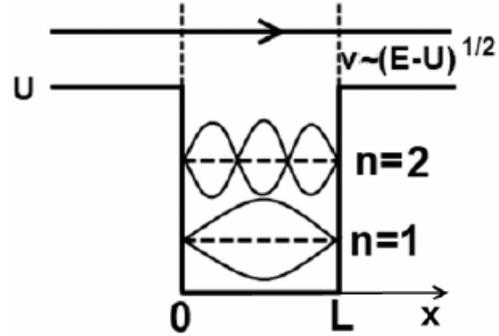

**Fig. 1.** *First two odd harmonics of standing acoustic waves in iodine air filter with length of L.*

The magnitude of the internal air stream velocity *V* is much smaller than the magnitude of the sound velocity in the air *s*, but it is important for the physical mechanism of acoustic oscillations generation in the air, which will be considered below.

The wave equation for the pressure oscillations in the air can be written as

$$\Delta P = \left(\frac{1}{s^2}\right)\frac{\partial^2 P}{\partial t^2}, \qquad (1)$$

where $P(x, y, z, t)$ is the pressure in the sound wave; $\Delta = \frac{\partial^2}{\partial x^2} + \frac{\partial^2}{\partial y^2} + \frac{\partial^2}{\partial z^2}$ is the *Laplace* operator; *t* the time. The sound velocity is defined by the dependence of the air density $\rho$ on the pressure $s^2 = (\partial P/\partial \rho)_S$ at the constant entropy *S*. It is visible that the magnitude of sound velocity in the granular filtering medium with the cylindrical coal granules should not be different from the magnitude of sound velocity in the free space, assuming that the internal air volumes of pores in the cylindrical coal granules are small in comparison with the free air volume of cavities between the cylindrical coal granules in the *IAF*. At the same time, it makes sense to explain that the internal absorbent surface of a cylindrical coal granule significantly exceeds the geometrical surface of a cylindrical coal granule [22, 23]. Then, this additional internal air volume in the cylindrical coal granules does not make any influence on the acoustic oscillations at the change of air pressures in an absorber, hence it has no effect on the sound velocity in the *IAF* [24].

The oscillatory velocity of movement of the small dispersive coal dust particles in the air, $v$, where the air is considered to be a continuous media, is an important parameter

$$v = \left(\frac{1}{\rho}\right) grad \int P dt. \quad (2)$$

On the boundary between the free space and the granular filtering medium, there are the following boundary conditions: $P_1 = P_2$ and $v_1 = v_2$. Let us note that, in the case of the acoustic oscillations with the wave length, which is much bigger than the size of the cylindrical coal granules, $\lambda \gg d_{GR}$, the magnitude of acoustic wave velocity in the granular filtering medium with the cylindrical coal granules will not be different from the magnitude of acoustic wave velocity in the free space.

The harmonic one-dimensional acoustic wave field can be described by the *Helmholtz* equation

$$\Delta P + k^2 P = 0, \quad (3)$$

where $k = \omega/s$ is the wave vector, $\omega = 2\pi f$ is the angular frequency, $f$ the linear frequency. In the general case, the acoustic wave can be described as

$$P(x,t) \sim \exp[i(kx - \omega t)],$$

where $k = k_1 + ik_2$ is the complex wave vector; the imaginary part of complex wave vector describes the attenuation of acoustic wave in the granular filtering medium

$$P(x,t) \sim \exp(-\alpha x) \cdot \exp[i(k_1 x - \omega t)],$$

where $\alpha = k_2$ in [25]. The coefficient of acoustic wave absorption in the granular filtering medium has a big value in the case of the acoustic oscillations with the wavelength, which is comparable or less to the characteristic sizes of the cylindrical coal granules. The scattering of acoustic waves, having $\lambda \gg d_{GR}$, can be assumed to be small in the considered research problem, hence it will have no any influence on the results [25].

In the case of an absorber, which is open for the air propagation (the case of the *IAF*), the sound pressure at the absorber's boundary reduces to the zero: $P = 0$; and the air particles velocity reaches its maximum value: $v = v_{max}$, i.e. the standing acoustic waves in the air, which are characterized by the odd harmonics, can only be generated in the *IAF*.

As the standing acoustic waves are set by the boundary conditions at which there are the anti-nodes of the acoustic waves velocity $v$, and the zeros of the air pressure; then the anti-nodes of the air pressure $P$ occur to be shifted on the one quarter of wavelength in the direction toward the *IAF*'s core.

In the *IAF*, the spectrum of possible acoustic waves is described by the expression

$$P_n = P_0 \sin\left[\frac{(2n-1)\pi x}{L}\right] \exp\left[\frac{i(2n-1)\pi s t}{L}\right], \quad (4)$$

where $L$ is the length of an adsorber; $n = 1, 2, 3 \ldots$. The frequencies of possible harmonics are interconnected by the following relation: $f_n = (2n - 1) f_1$, where $f_1$ is the basic harmonic with the wavelength: $\lambda = 2L$.

Let us consider the physical mechanisms, which can actually result in a generation of the acoustic oscillations in the *IAF*. For this purpose, let us estimate the main parameters, characterizing the air – dust aerosol propagation and the possible oscillatory processes in the granular medium, using such parameters as the *Reynolds number (Re)* and *Strouhal number (Sh)*.

The absorbent cylindrical coal granules in the *IAF*, can be conditionally replaced with the spherical granules of same volume with the diameter $D$. Then, the *Reynolds number (Re)* is

$$Re = \frac{\rho V D}{\mu} = \frac{VD}{\nu},$$

where $\nu = \mu/\rho$ is the kinematic viscosity of air. Substituting such variables as the characteristic diameter of granules, $D \approx 2.5 \cdot 10^{-3}$ m; the air density at the room temperature and $\rho \approx 1.2$ kg/m$^3$; $\mu = 1.78 \cdot 10^{-5}$ Pa·s at the normal air pressure; the speed of air stream $V = 2$ m/sec, we will get that the *Reynolds number* is $Re \approx 3.4 \cdot 10^2$. This aerodynamic regime of the *IAF*'s operation is accompanied by the turbulent excitations appearance in the air stream, flowing between the absorbent granules in the *IAF*.

The *Strouhal number* connects the frequency of the air flow oscillations $f$; the diameter of a sample $d$, which is flown by the air stream; and the local air stream velocity $V_l$:

$$Sh = \frac{d f}{V_l}.$$

At the *Strouhal number*, $Sh \approx 0.2$, and at the *Reynolds number*, $Re > 10^2$ in [26], there will be the process of the vortices separation from the body, resulting in the generation of the air flow oscillations.

These physical mechanisms of the acoustic waves generation are well known [27]. The acoustic oscillations are subdivided on the two possible classifications: *1)* the acoustic oscillations, connected with the air flow around the bodies of the final sizes (*Aeolus tone*) [28], and *2)* the acoustic oscillations, connected with the volumetric resonant structures in the form of the cavities between the granules in the *IAF* as in the researched case (*Helmholtz* resonator). In the first case, the characteristic resonant frequency of acoustic oscillations can be determined as

$$f = \frac{Sh \, V_l}{d}.$$

In the second case, the resonant frequency of acoustic oscillations is equal to

$$f = \frac{V_{res}}{d_{res}}$$

where $V_{res}$ is the specific velocity of the air flow in the *Helmholtz* resonator and $d_{res}$ the characteristic linear dimension of the free local volume between the granules. This frequency of *Helmholtz* resonator is close to the frequency of the *Strouhal* oscillations.

Thus, a part of the air stream's energy dissipates in the forced acoustic oscillations, exciting the standing acoustic waves in the *IAF*. In the general case, the acoustic oscillations, which are generated near to the various granules, can be synchronized [29, 30].

Let us notice that the wavelengths of acoustic waves at the specified frequencies significantly exceed the radiator's dimensions. Therefore, the radiation by the quasi-dot radiation source is almost isotropic and the radiation's intensity is small. However, all the granules can excite the acoustic oscillations in the *IAF*, hence it is necessary to take to the consideration the boundary conditions of all the radiators to characterize the radiation properly.

As it is known [25, 31], at the sufficient acoustic signal power, the standing acoustic waves can make the weighting influence on the small dispersive coal dust particles, even, on the large enough particles, in the gravitational field. This effect is based on the viscous capture of the small dispersive coal dust particles by the air streams, oscillating with the characteristic velocity, $\upsilon$ (eq. 2). Such viscous capture is important for the particles of the nano- and micro- sizes especially. The other ponderomotive effects, which are caused by the gradients of the air pressure and the density of kinetic energy, $W \approx \rho \upsilon^2$, at the action by the acoustic wave, can also appear [25]. The arising force influences on the larger coal dust particles mainly

$$F_n \sim W k^4 d^6,$$

where $W$ is the average density of the acoustic waves energy, $d$ is the diameter of a particle. The attraction forces, which facilitate the structurization of particles in the bigger structures, can also appear between the neighboring particles in the field of acoustic oscillations. Thus, the small dispersive coal dust particles collect in the regions of the anti-nodes of acoustic waves in the *IAF*. Taking to the consideration the influence by the first two odd harmonics of the acoustics oscillations, it makes sense to explain that the resulting spatial distribution of the small dispersive coal dust particles in the *IAF* will be equal to the sum of the two distributions in the *IAF*: *1)* the distribution, obtained at the actions by the two odd harmonics of the acoustics oscillations, and *2)* the distribution, obtained at an action by the constant air stream flow velocity $V$.

The difference of the air pressures, $\delta P$, at the input and output ports in the *IAF* is a forcing force, which creates the air stream flow in the *IAF*. The radiation of acoustic oscillations, generated by the air stream, should lead to the situation, when the variable, $\delta P$, is only responsible for the presence of the air stream flow $I(\delta P)$, resulting in the turbulent effects origination and some other accompanying acoustic phenomena appearance. It is a well known fact that this dependence between the forcing force and the created air stream $I$ is linear in the case of both the thermal-conductivity and the electrical-conductivity, when the physical mechanisms of energy dissipation are only connected with the scattering of quasi-particles at the inter-nuclear level in the crystals. The same is true in the case of the cylindrical tube with the diameter of $d$, which has no special geometric shape features and in which there are no turbulent air stream flow excitations

$$I = \frac{\sigma \, \delta P}{L},$$

where $\sigma = \frac{\pi d^4}{128 \nu}$ [26]. In the experiment with the *IAF*, the appearance of a new channel of dissipation results in the following dependence

$$I = \frac{\sigma S_{\square} (\delta P)^{2/3}}{L},$$

where $\sigma$ is the conditional conductivity of medium; $S_{\square}$ is the area of cross-section; $L$ is the device length. Such deviation from the linearity confirms an essential role by the nonlinear phenomena, which is connected with the non-homogeneity of the granular filtering medium, which causes a subsequent disappearance of the laminar properties of the air stream during its propagation in the *IAF*.

The presence of the forced standing acoustic wave oscillations leads to the maximal periodic change of the air pressures in the regions of the anti-nodes of the acoustic waves. This has to have an influence on the process of saturation of the small dispersive coal dust granules and the cylindrical coal granules by the adsorbed radioactive chemical elements and their isotopes, which are present in the air.

Let's notice that such oscillatory influence can result in a significant enhancement of the diffusional transposition of the small dispersive coal dust particles, changing the effective coefficient of diffusion on the orders of magnitude (see, for example, [32, 33]). We cannot clearly state that this physical mechanism is fully realized in this case, but its influence is quite possible, considering the fact that there are the anti-nodes of acoustic waves in the positions of maximums of accumulation of the adsorbed chemical elements and their isotopes as found by the gamma resonance spectroscopy method [2]. This effect must quite strongly appear in the case of the small particles, for example: in the small dispersive coal dust particles masses, because of their small sizes in comparison with the cylindrical coal granules; in view of the fact that the characteristic filling time of such small particles by the adsorbed chemical elements and their isotopes at the diffusion process is in the $(d_{GR}/d_{DUST})^2$ times faster than in the case of granules. As shown in [3], the specific mass saturation of the small dispersive coal dust particle by the adsorbed chemical elements and their isotopes is in a few times bigger, comparing to the cylindrical coal granules. This fact confirms the expressed suppositions about the role by the acoustic oscillations.

Also, it is necessary to consider the influence by the continuous air stream flow on the distributions of both the small dispersive coal dust particles and the adsorbed chemical elements and their isotopes, which can result in the possible shifts of their concentration density maxima along the length of an absorber, including the possible substitutions of one absorbed radioactive chemical elements by others, depending on the different values of their absorption energies in the *IAF*.

## Discussion on research results: comparison of theoretical calculations with experimental results

In the *IAF's* model of the type of *AU-1500*, the *IAF's* tube was made of the stainless steel, which can sustain the air stream flow up to *1500 m$^3$/hour*, with the unlimited operational time term. In the *AU-1500 IAF*, the granular filtering medium with the cylindrical coal granules of the type of *CKT-3*, which have the perfect absorption properties as far as all the radioactive *Iodine* isotopes, including the radioactive isotope $^{131}I$ with the decay period of ~ *8.14 days*, is concerned. The cylindrical coal granules represent the bodies of the cylindrical form with the average diameter *d*, which is close to *1.8 ÷ 2 mm*, and the average length *l ≈ 3.2 mm*. The length of the layer of the granular filtering medium with the cylindrical coal granules, through which the air is filtered, is *L ~ 0.3 m*, and the diameter of device is approximately ~ *1 m*. The internal structure of the cylindrical coal granules represents the complex quasi-fractal medium with the empty pores of various diameters, ranging from the meso-scope sizes to the nano-scope sizes, which are interconnected by the air channels with various cross-sections sizes. The internal effective adsorbing surface of such structure is big enough and it equals to a few hundreds of *m$^2$/g*, which results in the big absorption activity by the cylindrical coal granules in relation to a wide spectrum of chemical elements.

In our case with the averaged sizes of both the cylindrical coal granule and the free volume, the frequencies of the forcing acoustic oscillations are in the range of frequencies from *500 Hz* to *1500 Hz*, hence these frequencies coincide with the possible frequencies of the first and second odd harmonics of a main tone of the air density acoustic oscillation.

The ordered distribution of the small dispersive coal dust particles in the *IAF* [17]; and the similar distributions of the adsorbed radioactive chemical elements and their isotopes, correlated with the maxima of the small dispersive coal dust particles masses concentration density in the *IAF* [19]; were reported early. It was observed that there is an abnormal sharp increase of aerodynamic resistance and the sharp emissions of radionuclides in the *IAF* at some time moment during the *IAF's* operation. All these physical features can be explained, considering the influences by the standing acoustic waves, generated during the air stream scattering on the non-homogeneous granular filtering medium in the absorber in the *IAF*, on the processes of transposition of the small dispersive coal dust particles and the adsorbed chemical elements and their isotopes in the *IAF*.

Let's notice that the distribution of the accumulated small dispersive coal dust particles masses in the *IAF* [17], and the distribution of some absorbed radioactive chemical elements and their isotopes, measured by the gamma-activation spectroscopy method [19], are correlated as shown in the case radioactive *Strontium* ($^{88}Sr$) in Fig. 2.

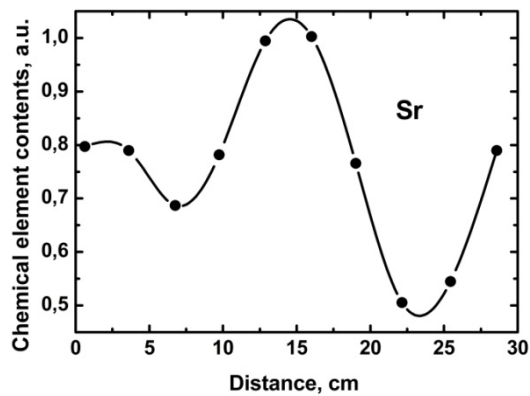

*Fig. 2. Characteristic dependence of relative distribution of radioactive $^{88}Sr$ along length of absorber with cylindrical coal granules in iodine air filter of type of AU-1500.*

We could come to the conclusion that the distributions [17, 19] are well correlated, if we would assume that all the three maxima appear as a result of the influence by the forcing standing acoustic oscillations on the distribution of the accumulated small dispersive coal dust particles masses and the absorbed radioactive chemical elements and their isotopes in the *IAF*. The physical forces, which act on the small dispersive coal dust particles in the oscillating air medium, can weigh the small dispersive coal dust particles in the field of the gravitation force in the vertical *IAF* in the regions, where there are the anti-nodes of the acoustic waves in the *IAF*. This is a main reason of the maxima origination in the distribution the small dispersive coal dust particles masses in regions of the anti-nodes of acoustic waves in the *IAF*.

In the very same regions, there are the intensive penetration and interchange of air in the volumes of internal pores in the cylindrical coal granules of the absorbent of the type of *CKT-3*. The same cyclic process is realized inside the pores of the small dispersive coal dust particles, structured in the particles clots in the absorber in the *IAF*.

Let us conclude by saying that, in the presented research, the comprehensive analysis on the influences by the forcing standing acoustic oscillations, generated by the air stream flow in the *IAF*, on the distribution of the small dispersive coal dust particles and the radioactive chemical elements and their isotopes in the absorber in the *IAF* during the long-term *IAF's* operation at the *NPP* is completed for the first time. This analysis allowed us to precisely characterize the *IAF* parameters during the process of its operation at the *NPP*, and to make the innovative propositions on the necessary technical modifications with the purpose to improve the *IAF's* technical characteristics and increase the *IAF's* operational time at the nuclear power plant.

## Conclusion

The physical features of distribution of the small dispersive coal dust particles and the adsorbed radioactive chemical elements and their isotopes in the absorber with the granular filtering medium with the cylindrical coal granules were researched in the case of the intensive air-dust aerosol stream flow through the *IAF*. It was shown that, at the certain aerodynamic conditions in the *IAF*, the generation of the acoustic oscillations is possible. It was found that the acoustic oscillations generation results in an appearance of the standing acoustic waves of the pressure (density) in the *IAF*. In the case of the intensive blow of the air-dust aerosol, it was demonstrated that the standing acoustic waves have some strong influences on both: *1)* the dynamics of small dispersive coal dust particles movement and their accumulation in the *IAF*; *2)* the over-saturation of the cylindrical coal granules by the adsorbed radioactive chemical elements and their isotopes in the regions, where the anti-nodes of the acoustic waves are positioned, in the absorber in the *IAF*. Finally, we completed the comparative analysis of the theoretical calculations with the experimental results, obtained for the cases of: *1)* the experimental aerodynamic modeling of physical processes of the absorbed radioactive chemical elements and their isotopes distribution in the *IAF*; and *2)* the gamma activation spectroscopy analysis of the absorbed radioactive chemical elements and their isotopes distribution in the *IAF*. We made the innovative propositions on the necessary technical modifications with the purpose to improve the *IAF's* technical characteristics and increase the *IAF's* operational time at the nuclear power plant, going from the completed precise characterization of the *IAF's* parameters at the *IAF's* long term operation at the *NPP*.

Authors are very grateful to a group of leading scientists from the *National Academy of Sciences in Ukraine* (*NASU*) for the numerous encouraging scientific discussions on the reported theoretical and experimental research results.

This innovative research is completed in the frames of the nuclear science and technology fundamental research program, facilitating the environment protection from the radioactive contamination, at the *National Scientific Centre Kharkov Institute of Physics and Technology* (*NSC KIPT*) in Kharkov in Ukraine. The research is funded by the *National Academy of Sciences in Ukraine* (*NASU*).

This research article was published in the *Problems of Atomic Science and Technology* (*VANT*) in [34] in 2013.

*E-mail: ledenyov@kipt.kharkov.ua

___________